\documentclass[runningheads]{llncs}
\usepackage{booktabs}
\usepackage{xcolor}

\usepackage{multirow,multicol}
\usepackage{cite}
\usepackage{amsmath} 
\usepackage[linesnumbered,ruled,vlined]{algorithm2e}
\usepackage[T1]{fontenc}
\usepackage{graphicx,verbatim}
\begin{document}
\newcommand{\EAc}[1]{\textcolor{cyan}{#1}}
\newcommand{\EA}[1]{\textcolor{magenta}{#1}}
\newcommand{\AK}[1]{\textcolor{red}{#1}}
\title{Boundary-Emphasized Weight Maps for Distal Airway Segmentation}
%
\begin{comment}  %% Removed for anonymized MICCAI 2025 submission
\author{First Author\inst{1}\orcidID{0000-1111-2222-3333} \and
Second Author\inst{2,3}\orcidID{1111-2222-3333-4444} \and
Third Author\inst{3}\orcidID{2222--3333-4444-5555}}
%
\authorrunning{F. Author et al.}
\institute{Princeton University, Princeton NJ 08544, USA \and
Springer Heidelberg, Tiergartenstr. 17, 69121 Heidelberg, Germany
\email{lncs@springer.com}\\
\url{http://www.springer.com/gp/computer-science/lncs} \and
ABC Institute, Rupert-Karls-University Heidelberg, Heidelberg, Germany\\
\email{\{abc,lncs\}@uni-heidelberg.de}}

\end{comment}

\author{Ali Keshavarzi and Elsa Angelini}  %% Added for anonymized MICCAI 2025 submission
\authorrunning{Anonymized Author et al.}
\institute{LTCI, Telecom Paris, Institut Polytechnique de Paris, Palaiseau, France \\
    \email{ali.keshavarzi@ip-paris.fr}}

\maketitle              % typeset the header of the contribution
\begin{abstract}
Automated airway segmentation from lung CT scans is vital for diagnosing and monitoring pulmonary diseases. Despite advancements, challenges like leakage, breakage, and class imbalance persist, particularly in capturing small airways and preserving topology. We propose the Boundary-Emphasized Loss (BEL), which enhances boundary preservation using a boundary-based weight map and an adaptive weight refinement strategy. Unlike centerline-based approaches, BEL prioritizes boundary voxels to reduce misclassification, improve topology, and enhance structural consistency, especially on distal airway branches. Evaluated on ATM22 and AIIB23, BEL outperforms baseline loss functions, achieving higher topology-related metrics and comparable overall-based measures. Qualitative results further highlight BEL’s ability to capture fine anatomical details and reduce segmentation errors, particularly in small airways. These findings establish BEL as a promising solution for accurate and topology-enhancing airway segmentation in medical imaging.


\end{abstract}

\section{Introduction}
\label{sec:introduction}
Airway segmentation from lung CT scans is essential for diagnosing and monitoring pulmonary diseases such as bronchiectasis, and chronic obstructive pulmonary disease. However, manual segmentation is time-consuming, labor intensive, and prone to user variability.

Despite advancements in deep learning (DL)-based airway segmentation, key challenges remain, including \textit{leakage}, \textit{breakage}, and gradient-related errors like \textit{dilation} and \textit{erosion}\cite{zheng21_wingsnet}. Leakage results in over-segmentation due to airway lumen intensity variability\cite{charbonnier17_leakage}, while breakage disrupts connectivity, particularly in finer branches~\cite{zhang23_atm22, zhang24_surrloss}. Gradient erosion shrinks small airways, whereas dilation overextends predictions, degrading segmentation accuracy~\cite{zheng21_wingsnet}.

% GOAL class-wise imbalance
Another major challenge is class imbalance, as airway voxels constitute only 2–3\% of a CT scan~\footnote{Statistics derived from ATM22~\cite{zhang23_atm22} dataset.}. This imbalance is exacerbated by airway diameter variation, with the trachea being significantly larger than distal airways, biasing overlap-based loss functions such as Dice~\cite{milletari16_dice}, Tversky~\cite{deniz17_tversky}, and Focal Loss~\cite{wang18_dice_focal, nabila19_tversky_focal}.
This imbalance also affects topological metrics such as Detected Length Rate and Detected Branch Rate, used for assessing airway connectivity.

% GOAL --> use of prior knowledge in airway segmentation
A promising strategy for improving airway segmentation involves integrating prior anatomical knowledge, often using centerline-based distance maps~\cite{wang19_radial_distance, zheng21_wingsnet, ke23_tversky}. However, these methods depend on skeletonization accuracy. Errors in skeleton extraction propagate to weight maps, reducing reliability, especially near the airway lumen. Additionally, skeleton-based weight maps often exacerbate leakage and degrade precision.

In this paper, we introduce a novel \textbf{Boundary-Emphasized Loss (BEL)}, which prioritizes airway boundaries through a custom weight map that emphasizes edge voxels. Additionally, we propose an \textbf{adaptive weight refinement module} that dynamically adjusts weight maps at breakage locations, iteratively improving segmentation continuity. Our method is evaluated against conventional overlap-based and centerline-based losses to assess its effectiveness in both overlap and topological metrics. Our key contributions are:
\begin{itemize}
    \item \textbf{Boundary-Emphasized Loss (BEL):} A novel loss function leveraging airway edge-focused weight maps.
    \item \textbf{Adaptive Weight Refinement:} A dynamic strategy that iteratively adjusts weight maps during training to improve segmentation in challenging airway regions.
    \item \textbf{Comprehensive Evaluation:} Extensive comparisons against overlap-wise and topological loss functions to assess performance across airway scales.
\end{itemize}

\section{Related Works}
\subsection{Airway Segmentation}
Traditional image processing methods~\cite{tschirren05_trad, fabijanska09_trad, rikxoort09_trad} have been largely outperformed by deep learning (DL) models in airway segmentation, particularly CNN-based architectures like UNet~\cite{Ronneberger15_unet}, nnUNet~\cite{Isensee21_nnunet}, and ResUNet~\cite{Ilkay19_resunet}. More specialized networks, including AirwayNet~\cite{qin19_airwaynet}, WingsNet~\cite{zheng21_wingsnet}, and TACNet~\cite{cheng21_tacnet}, further refine feature extraction, enhancing segmentation performance~\cite{garcia21_airseg, zhang24_surrloss}. While transformer-based models have advanced natural and medical image segmentation~\cite{dosovitskiy21_vit, liu21_swint, hatamizadeh22_unetr, hatamizadeh21_swin}, they remain less effective than CNN-based encoder-decoder models for volumetric segmentation due to high computational demands and limited local context~\cite{zhang23_atm22}. Airway variability and the limitations of overlap-based loss functions continue to drive research in this field~\cite{zheng21_wingsnet}. 

% Multi-scale issue

\subsection{Loss Functions}
Accurate airway segmentation requires loss functions that balance overlap-based accuracy with topological preservation. Dice loss is widely used in this domain due to its effectiveness in handling class imbalance and sparse foreground structures~\cite{milletari16_dice, garcia21_airseg}. 
To improve performance, various studies have explored hybrid loss functions. Zhang et al.~\cite{zhang20_dicel2} combined Dice with L2 loss in a cascaded 2D+3D model, enhancing segmentation accuracy in pathological CT scans, particularly for peripheral airways. Juarez et al.~\cite{garcia18_bce_dice} compared weighted binary cross-entropy (wBCE) and Dice loss, using dynamic weights to mitigate class imbalance between lung parenchyma and airway voxels. Similarly, Zhang et al.~\cite{zhang21_fda} integrated Dice with Focal loss, yielding improvements in bronchial segmentation under noisy conditions. Notably, the majority of ATM22 airway segmentation challenge participants~\cite{zhang23_atm22} employed Dice-based loss functions. 

While Dice loss and its variants enhance segmentation precision and overlap metrics, Tversky loss~\cite{deniz17_tversky} addresses inter-class imbalance and improves model calibration for False Negatives~\cite{hu2024_attnet}. However, even advanced loss functions struggle with discontinuities in distal airways due to extreme class imbalance~\cite{wang19_radial_distance, zheng21_wingsnet, ke23_tversky}.
Despite efforts to distinguish airway branches by diameter~\cite{wang19_radial_distance, zheng21_wingsnet, ke23_tversky}, models still struggle with local discontinuity, particularly in distal airways where class imbalance is severe. Wang et al.\cite{wang19_radial_distance} introduced Radial Distance Loss (RDL), prioritizing centerline voxels with radially decreasing weights to enhance fine airway segmentation. Zheng et al.\cite{zheng21_wingsnet} extended this with the General Union Loss (GUL), using size-adaptive weight maps, while Ke et al.~\cite{ke23_tversky} refined radial weighting with segmentation predictions. Nonetheless, consistent segmentation of distal airways remains challenging, necessitating improved loss formulations.

\section{Methodology}
\label{sec:methods}

\subsection{Boundary-Emphasized Loss (BELoss)}
Our loss function extends the Root Tversky loss~\cite{nabila19_tversky_focal} by integrating airway boundary priors. Unlike centerline-based methods~\cite{wang19_radial_distance, zheng21_wingsnet, ke23_tversky}, we use boundary emphasized priors and a soft breakage detection module to mitigate disconnections. 

\textbf{Intuition.} Distance to boundaries better reflect the shape of the airway and is less noisy than distance to centerline. During training a dynamic mechanism to specifically focus on preventing breakages can be inserted as an additional weight map.

\textbf{BEL:}
Following the formulation of the Root Tversky  and GUL losses, we define BEL as:
\begin{equation}
\text{BELoss}  = 1 - \frac{\sum_{i=1}^{N} w_i \, p_i^{r} \, g_i}{\sum_{i=1}^{N} w_i \left ( \alpha \, p_i \;+\; \beta \, g_i \right )}, 
\label{eq:beloss}
\end{equation}

\noindent where \(p_i\) and \(g_i\) denote the predicted probability and ground truth label for voxel \(i\), respectively, and \(\alpha,\beta\) are balancing parameters. The exponent \(r \in (0,1)\) introduces some non-linearity to emphasize voxels with lower predicted probabilities. \(w_i\) is our boundary- and breakage-aware weight, described next.

\textbf{Weight Map \(\mathbf{w_i}\):}  
Background voxels (\(g_i=0\)) receive a constant weight of 1, while airway (foreground) voxels are weighted based on their distance to the \emph{airway boundary} and potential breakages:
\begin{equation}
w_i \;=\; \biggl( 1 - \mu \,\Bigl(\tfrac{d_i}{d_{\max}}\Bigr)^{\gamma} \biggr) 
\;\times\;\Bigl(1 + \theta \,B_i\Bigr),
\label{eq:beloss_w}
\end{equation}

\noindent where \(d_i\) is the distance from voxel \(i\) to the nearest boundary (computed by \texttt{distance\_transform\_edt} from \texttt{scipy.ndimage}), and \(d_{\max}\) is the maximal boundary distance value within the tree.
The \(\gamma\) parameter controls the weight decay and \(\mu\) serves as a boundary emphasis scaling factor. The term \(B_i\) represents a \emph{soft breakage map}, identifying under-segmented regions without discrete skeletonization (see below), while \(\theta\) modulates its influence.

\textbf{Boundary Extraction.} 	
To extract airway boundaries, we apply a function named \texttt{binary\_erosion} from \texttt{scipy.ndimage} on the airway mask, then identify boundary voxels by computing the difference between the original mask and its eroded version using \texttt{np.logical\_and}. This process isolates the outermost airway surface, forming the basis for boundary-aware weighting. 

\textbf{Soft Breakage Detection.}
We adopt a \emph{soft skeletonization} strategy inspired by clDice~\cite{cldice} to generate continuous skeleton representations for both the ground truth (\(g\)) and the prediction (\(p\)) during training. Instead of relying on discrete morphological operations, we use \texttt{soft\_erode} and \texttt{soft\_dilate} detailed in~\cite{cldice}, which apply differentiable max-pooling functions to iteratively refine airway structures while preserving differentiability. Given 3D binary masks \(g\) and \(p\), we obtain soft skeleton maps:  

\[
S_{GT} = \texttt{soft\_skel}(g), \quad S_{P} = \texttt{soft\_skel}(p),
\]
where each \(\texttt{soft\_skel}\) call yields a \emph{continuous} skeleton map indicating the core or centerline strength of the respective volume. We then define a breakage map:
\begin{equation}
B_i \;=\; \max\bigl(0,\; S_{GT}(i) - S_{P}(i)\bigr).
\label{eq:soft_breakage}
\end{equation}
Here, \(B_i \in [0,1]\) measures how much skeleton from the ground truth is missing in the prediction at voxel \(i\). If \(p\) accurately captures the airway structure, then \(S_{P} \approx S_{GT}\) and \(B_i\approx0\). In contrast, insufficient coverage of thin or distal branches increases \(B_i\), reflecting potential breakages. Finally, \(B_i\) is incorporated into the boundary-emphasized weight \(w_i\) (see Eq.~\eqref{eq:beloss_w}) via the term \(\bigl(1 + \theta\,B_i\bigr)\), thus boosting emphasis on regions most prone to topological disconnections. The \emph{soft} definition avoids the non-differentiable set operations associated with discrete skeletonization while still effectively highlighting connectivity gaps. 

\section{Experiments}

\subsection{Datasets}
We use two open-access cohorts with pixel-level annotations. 
The \textbf{ATM22}~\cite{zhang23_atm22} dataset comprises 500 CT scans from various scanners. It is officially split into 300 scans for training, 150 for testing, and 50 for validation. Each scan contains between 84 and 1,125 axial slices of size \(512 \times 512\) voxels, with in-plane voxel sizes ranging from 0.51mm to 0.91mm  and 0.5mm to 5.0mm slice thickness. Ground-truth annotations include the trachea, main bronchi, lobar bronchi, and distal segmental bronchi.
The \textbf{AIIB23}~\cite{aiib23_ds} dataset consists of 285 CT scans, among which 235 from fibrotic lung disease patients and 50 from COVID-19 cases. Each scan contains between 146 and 947 slices of size \(512 \times 512\) or \(768 \times 768\) voxels, with in-plane voxel sizes ranging from 0.41mm to 0.92mm and 0.39mm to 2.0mm in slice thickness. Ground-truth annotations include the same airway regions as in the ATM22 dataset.
Datasets were split into 80\% training and 20\% validation/evaluation set at scan level.
To minimize sampling bias, we used 5-fold cross-validation for robust evaluation. For ATM22, $N_{\textit{ATM22}} = 299$ publicly available scans\footnote{One Case, excluded by ATM22 challenge organizers.} (239:60) were used for training and validation/evaluation, except in the last fold (240:59). For AIIB23, $N_{\textit{AIIB}} = 120$ publicly available scans (96:24) were used for training and validation/evaluation. 

\subsection{Preprocessing}
For training, Hounsfield Unit (HU) intensity values are clipped to \([-1000,600]\) and rescaled to \([0,1]\). Lungs are segmented using the open-source LungMask~\cite{Hofmanninger2020} method. The lung region is extracted using the lung mask and extended vertically to include the ground-truth trachea. 
No resampling was performed.
Random cropping extracts \(256 \times 256 \times 256\) patches for both training and validation, ensuring that the center is inside the lung, to work with variable lung crops.

For evaluation (inference on full field-view of validation scans), preprocessing is identical except for the following modifications. The lung region is cropped using the lung mask, and extended vertically by 50 voxels on top of the lung (sufficient for all slice thicknesses).
Instead of random cropping, 3D patches are extracted using a sliding window with a 25\% overlap to cover the lung region.

\subsection{Implementation Details and Hyper-Parameters}

\textbf{Deep Architecture.} We use a 3D Attention U-Net architecture with five layers, progressively increasing dimensions \([16, 32, 64, 128, 256]\). Both encoder and decoder paths use \(3 \times 3 \times 3\) kernels with a stride of 2 for downsampling/upsampling. Separate models were trained for the ATM22 and AIIB23 datasets.
We tested 4 models with different loss functions: (1) Dice: using only the Dice loss function, (2) Tversky: using Eq. \ref{eq:beloss} and setting $w_i$ =1 everywhere and $r$=1 , (3) GUL: with Eq. \ref{eq:beloss} and $d_i$ in Eq. \ref{eq:beloss_w} based on distance to centerline (implementation provided by the GUL authors) and with $B_i$ set to 0, and (4) BEL: our proposed formulation with $d_i$ in Eq. \ref{eq:beloss_w} based on distance to boundaries and with $B_i$ computed dynamically during training. 

\noindent \textbf{Training parameters.} The model is trained with a learning rate of \(1 \times 10^{-3}\), batch size 3, and the Adam optimizer, applying a weight decay of \(1 \times 10^{-5}\) to the model parameters. Training runs for a maximum of 350 epochs (or 20 hours). A \texttt{ReduceLROnPlateau} scheduler reduces the learning rate by 0.1 if the validation loss does not improve for 30 epochs.  

\noindent \textbf{Loss hyperparameters.}
For GUL and BEL, following\cite{zheng21_wingsnet} we set terms in Eq. \ref{eq:beloss} and Eq. \ref{eq:beloss_w}, $\alpha$=0.2, $\beta$=1-$\alpha$ and $\mu$ = $\frac{1-2\alpha}{1-\alpha} = 0.75$. Given the sensitivity of the losses to the other two hyperparameters (exponents $r$ and $\gamma$) both BEL and GUL were optimised across an exhaustive search over the following ranges of values: \(\gamma\) = \([0.4, 0.6, 0.8, 1]\), and \(r\) = \([0.5, 0.7]\). For BEL, we used $\theta$=0.05.
The best evaluation results are reported. 

\noindent\textbf{Hardware resources.} The preprocessing and training of the Attention U-Net model were implemented using the Monai framework~\cite{cardoso2022monaiopensourceframeworkdeep} and PyTorch Lightning~\cite{Falcon_PyTorch_Lightning_2019}. Training was conducted on two NVIDIA A100-SXM4-80GB GPUs.

\section{Results and Discussions}
\subsection{Quantitative Results}
We report in Table~\ref{tbg:quant_lcc} quantitative segmentation results using the following metrics: IoU, Detected Length Rate (DLR), Detected Branch Rate (DBR), Precision, Leakage and Airway Miss Ratio (AMR). The details of these metrics are provided in~\cite{aiib23_ds}.

BEL reaches superior performance than Dice, Tversky, and GUL models for the DLR, DBR, and 1-AMR metrics on both AIIB23 and ATM22. 
In AIIB23, BEL outperforms GUL by \textbf{2.83\%} (DLR), \textbf{3.86\%} (DBR), and \textbf{0.58\%} (1-AMR). Similarly, in ATM22, BEL surpasses GUL by \textbf{5.48\%} (DLR), \textbf{7.71\%} (DBR), and \textbf{1.88\%} (1-AMR). Notably, GUL shows higher standard deviations in DLR and DBR across both datasets, with particularly high values in ATM22 (7.5\% and 9.9\%, respectively), indicating inconsistent performance. In contrast, BEL maintains significantly lower standard deviations (~0.8\%) for ATM22 and ~1.8\% for AIIB23, demonstrating greater stability in segmentation quality.

We report in Table~\ref{tbg:quant_lcc_small} the IoU\textsubscript{s}, DLR\textsubscript{s} and DBR\textsubscript{s} which correspond to the same metrics as in Table~\ref{tbg:quant_lcc} but  measured only in small-branches (airways inside the lung, excluding the trachea and main bronchi). All metrics decrease when measured only on small branches. 
In AIIB23, BEL achieves the highest DLR\textsubscript{s} and DBR\textsubscript{s}, surpassing GUL by \textbf{3.2\%} and \textbf{3.85\%}, reinforcing its ability to maintain airway continuity at finer scales. 
The Dice model struggles the most on AIIB23, with DLR\textsubscript{s} and DBR\textsubscript{s} of \textbf{61.49\%} and \textbf{50.83\%}. On ATM22, the Dice model is the best in terms of IoU but not for the other metrics. 
%improves, reaching DLR\textsubscript{s} and DBR\textsubscript{s} of \textbf{81.19\%} and \textbf{70.92\%}, yet 
BEL remains the best for DLR and DBR, surpassing GUL by \textbf{6.16\%} in DLR\textsubscript{s} and \textbf{7.84\%} in DBR\textsubscript{s}. 
All methods lose 10 to 20\% in performance metrics when tested on the disease dataset (AIIB23) compared to the more healthy one (ATM22). %confirming its robustness across datasets.

While BEL excels in topological preservation, it slightly underperforms in overlap-based metrics (IoU) and voxel-level measures such as Precision and 1-Leakage. Specifically, BEL scores lower than GUL in IoU across both datasets and exhibits reduced precision compared to more naive losses. This performance gap may stem from missing ground-truth annotations on the smallest airways, hence counted as false-positive voxels by overlap metrics. 

\begin{table*}[h!]
\begin{center}
\begin{sc}
\begin{tabular}{c|c|c|c|c|c|c}
\specialrule{0.2em}{1pt}{1pt} \addlinespace[4pt]
\multirow{2}{*}{Loss} & IoU & DLR & DBR & Prec. & 1-Leak. & 1-AMR \\
                      & (\%) & (\%) & (\%) & (\%)  & (\%)   & (\%) \\
\specialrule{0.1em}{1pt}{1pt} \addlinespace[4pt]
\multicolumn{7}{c}{AIIB23} \\
\specialrule{0.1em}{1pt}{1pt} \addlinespace[4pt]
dice     & $84.13 \pm 3.7$ & $66.03 \pm 3.8$ & $56.51 \pm 3.5$ & $\mathbf{94.97 \pm 1.2}$ & $\mathbf{96.06 \pm 0.8}$ & $88.29 \pm 2.6$ \\
tversky  & $84.16 \pm 4.0$ & $75.47 \pm 6.6$ & $67.79 \pm 6.9$ & $92.49 \pm 1.6$ & $92.58 \pm 2.1$ & $90.31 \pm 5.2$ \\
$\text{gul}_{0.6}$ & $\mathbf{85.82 \pm 2.0}$ & $79.35 \pm 2.9$ & $71.84 \pm 3.3$ & $92.33 \pm 1.2$ & $92.11 \pm 1.4$ & $92.50 \pm 2.7$ \\
$\text{bel}_{0.8}$ & $84.80 \pm 3.1$ & $\mathbf{82.18 \pm 1.7}$ & $\mathbf{75.70 \pm 1.9}$ & $90.44 \pm 1.0$ & $90.10 \pm 1.5$ & $\mathbf{93.08 \pm 3.8}$ \\
\specialrule{0.1em}{1pt}{1pt} \addlinespace[4pt]
\multicolumn{7}{c}{ATM22} \\
\specialrule{0.1em}{1pt}{1pt} \addlinespace[4pt]
dice     & $\mathbf{86.13 \pm 0.5}$ & $83.54 \pm 1.5$ & $75.86 \pm 2.2$ & $\mathbf{92.06 \pm 0.6}$ & $\mathbf{91.60 \pm 0.7}$ & $93.19 \pm 0.6$ \\
tversky  & $83.43 \pm 1.6$ & $79.70 \pm 8.8$ & $71.34 \pm 12.3$ & $90.36 \pm 0.9$ & $89.81 \pm 1.1$ & $91.77 \pm 2.1$ \\
$\text{gul}_{0.6}$ & $85.61 \pm 1.0$ & $88.59 \pm 7.5$ & $84.04 \pm 9.9$ & $89.39 \pm 2.4$ & $88.15 \pm 3.2$ & $95.47 \pm 2.4$ \\
$\text{bel}_{0.6}$ & $84.57 \pm 1.5$ & $\mathbf{94.07 \pm 0.8}$ & $\mathbf{91.75 \pm 0.8}$ & $86.68 \pm 1.6$ & $84.47 \pm 2.3$ & $\mathbf{97.35 \pm 0.5}$ \\
\specialrule{0.2em}{1pt}{1pt}
\end{tabular}
\end{sc}
\end{center}
\caption{Quantitative segmentation quality metrics computed on the largest connected components for evaluation sets AIIB23 and ATM22. Mean and standard deviation are reported for a 5-fold cross-validation. Results are reported for 4 models. The subscript of the loss names represents the $\gamma$ value and $r=0.7$ for both GUL and BEL.
}
\label{tbg:quant_lcc}
\end{table*}

\begin{table*}[htb!]
\begin{center}
\begin{minipage}{0.49\textwidth}
\centering
\fontsize{8}{9}\selectfont
\begin{sc}
\begin{tabular}{c|c|c|c}
\specialrule{0.2em}{1pt}{1pt} \addlinespace[4pt]
\multirow{2}{*}{Loss} & $IoU_{s}$ & $DLR_{s}$ & $DBR_{s}$ \\
                      & (\%)      & (\%)      & (\%)      \\
\specialrule{0.1em}{1pt}{1pt} \addlinespace[4pt]
Dice    & $63.58 \pm 4.9$ & $61.49 \pm 4.3$ & $50.83 \pm 3.8$ \\
\cmidrule{1-4}
Tver. & $65.79 \pm 4.9$ & $72.17 \pm 7.6$ & $62.05 \pm 7.1$ \\
\cmidrule{1-4}
$\text{gul}_{0.6}$ & $\mathbf{68.42 \pm 2.7}$ & $76.52 \pm 3.0$ & $65.99 \pm 3.3$ \\
\cmidrule{1-4}
$\text{bel}_{0.8}$ & $68.10 \pm 3.1$ & $\mathbf{79.72 \pm 2.1}$ & $\mathbf{69.84 \pm 2.1}$ \\
\specialrule{0.2em}{1pt}{1pt}
\end{tabular}
\end{sc}
% \caption{AIIB23}
\label{tab:aiib}
\end{minipage}%
\hfill
\begin{minipage}{0.49\textwidth}
\centering
\fontsize{8}{9}\selectfont
\begin{sc}
\begin{tabular}{c|c|c|c}
\specialrule{0.2em}{1pt}{1pt} \addlinespace[4pt]
\multirow{2}{*}{Loss} & $IoU_{s}$ & $DLR_{s}$ & $DBR_{s}$ \\
                      & (\%)      & (\%)      & (\%)      \\
\specialrule{0.1em}{1pt}{1pt} \addlinespace[4pt]
Dice    & $\mathbf{76.48 \pm 0.8}$ & $81.19 \pm 1.6$ & $70.92 \pm 2.0$ \\
\cmidrule{1-4}
Tver. & $71.44 \pm 3.4$ & $77.38 \pm 9.6$ & $66.52 \pm 12.0$ \\
\cmidrule{1-4}
$\text{gul}_{0.6}$ & $74.41 \pm 3.6$ & $87.08 \pm 8.2$ & $78.92 \pm 10.1$ \\
\cmidrule{1-4}
$\text{bel}_{0.6}$ & $73.94 \pm 3.0$ & $\mathbf{93.24 \pm 0.9}$ & $\mathbf{86.76 \pm 0.8}$ \\
\specialrule{0.2em}{1pt}{1pt}
\end{tabular}
\end{sc}
% \caption{ATM22}
\label{tab:atm}
\end{minipage}
\end{center}
\caption{Quantitative segmentation quality metrics as in Table~\ref{tbg:quant_lcc}, measured on small-airways inside the lung, for the AIIB23 (left) and ATM22 (right) datasets.}
\label{tbg:quant_lcc_small}
\end{table*}

\subsection{Qualitative Results}
\begin{figure}[htb!]
    \centering
    \includegraphics[width=\columnwidth]{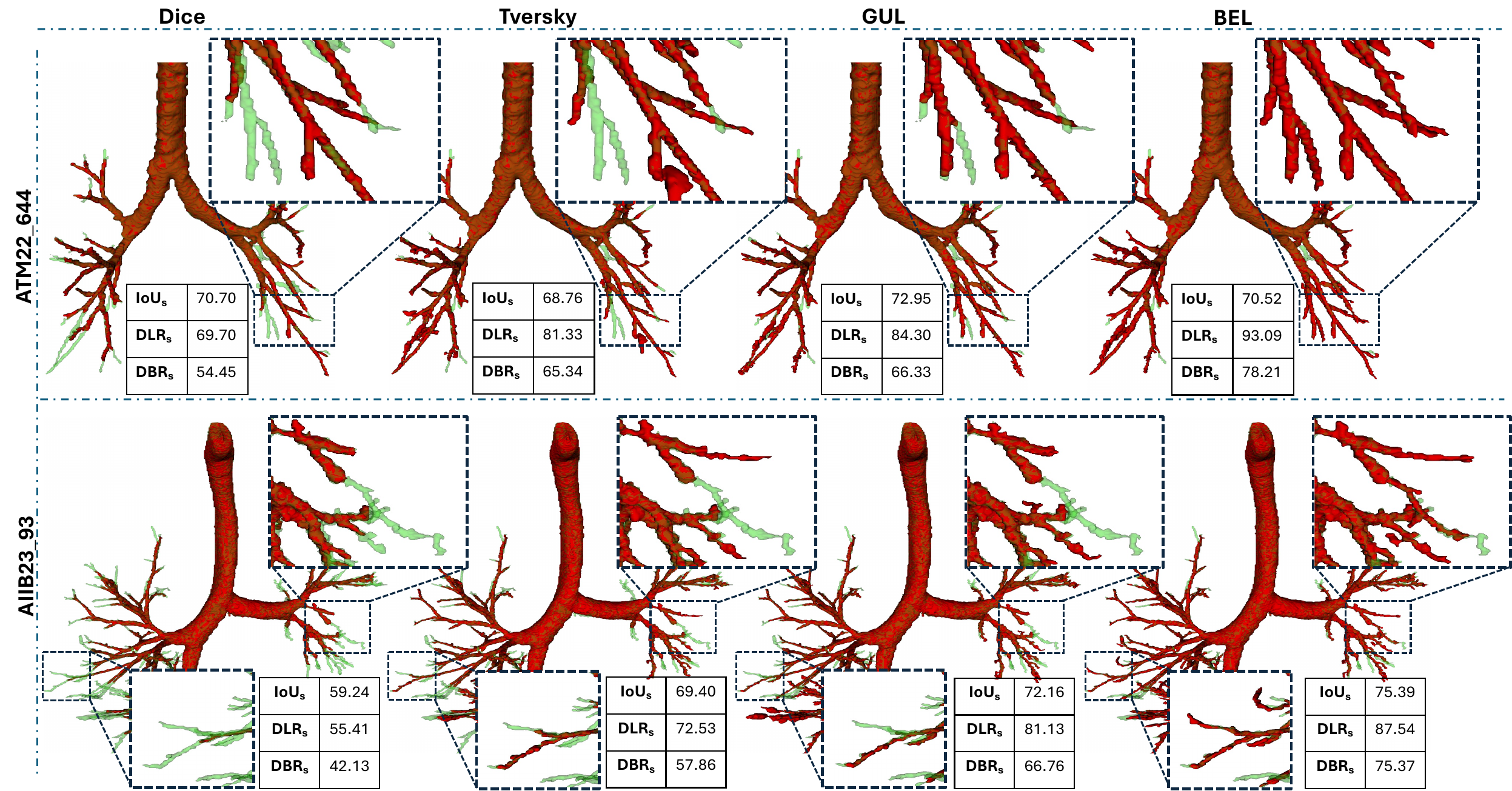}
    \caption{3D rendering of segmentation results on one case per dataset (ATM22 and AIIB23) along with small-airways performance metrics. Red=segmentation result, Green=ground-truth.}
    \label{fig:qualitative}
\end{figure}

We illustrate the quality of the compared segmentation models (BEL, GUL, Dice, and Tversky) on the overall airway tree and on small airways. In Figure~\ref{fig:qualitative} we provide 3D rendering of the largest connected component extracted from the segmentation mask generated by each model on one case per dataset (AIIB23 and ATM22). BEL excels in preserving finer airway structures, particularly in AIIB23, where Dice and Tversky often miss distal branches, and GUL struggles with consistency in smaller segments. In ATM22, BEL achieves superior structural integrity, especially in peripheral airways, outperforming other losses that tend to under-segment smaller branches.

\begin{figure}[htb!]
\centering
\includegraphics[width=\columnwidth]{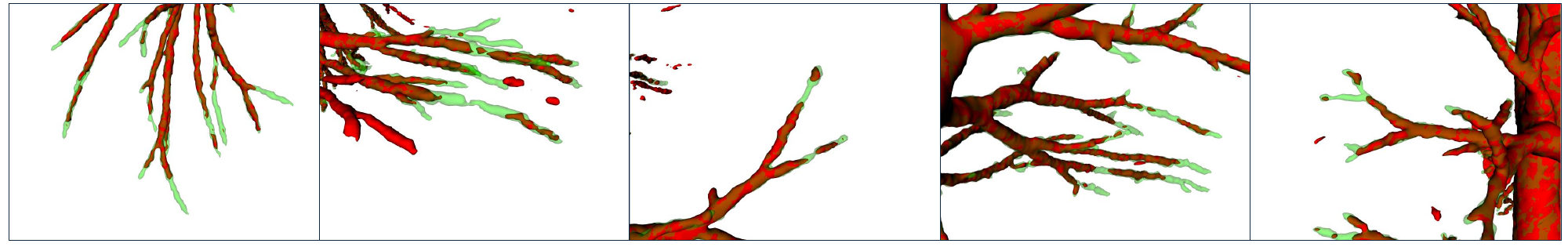}
\caption{3D rendering of BEL segmentation results with (green) and without ($B_i = 0$) (red) the proposed adaptive breakage weight map in the loss function during training.}
\label{fig:adaptive_wc}
\end{figure}

We illustrate in Figure~\ref{fig:adaptive_wc}  the impact of our proposed breakage weight map in our proposed BEL loss function for preventing breakages in smaller airways. 

\section{Conclusion}
We proposed a novel Boundary-Emphasized Loss (BEL) function which achieves superior airway tree segmentation on lung CT scans on two open-cohorts (AIIB23 and ATM22) associated with open Challenges. We compared internal evaluation results to 3 models using alternative loss functions (Dice, Tversky, and GUL) trained on a shared common CNN 3D Attention U-Net architecture. By emphasizing boundary voxels and incorporating adaptive weighting at breakage locations, BEL effectively reduces breakages, enhances small-branch detection, and preserves anatomical continuity. 
These findings confirm BEL’s highest robustness and accuracy on small airways. 
Evaluation on the external test sets of these Challenges remain to be run (on-going). We also consider benchmarking against alternative approaches such as learning to correct for missed airway parts from synthetic ground-truth degradations as in~\cite{CHEN2025103355}.

\begin{comment}  %% removed for anonymized MICCAI 2025 submission.
    
    % The following acknowledgement and disclaimer sections should be removed for the double-blind review process.  
    % If and when your paper is accepted, reinsert the acknowledgement and the disclaimer clause in your final camera-ready version.

\begin{credits}
\subsubsection{\ackname} A bold run-in heading in small font size at the end of the paper is
used for general acknowledgments, for example: This study was funded
by X (grant number Y).

\subsubsection{\discintname}
It is now necessary to declare any competing interests or to specifically
state that the authors have no competing interests. Please place the
statement with a bold run-in heading in small font size beneath the
(optional) acknowledgments\footnote{If EquinOCS, our proceedings submission
system, is used, then the disclaimer can be provided directly in the system.},
for example: The authors have no competing interests to declare that are
relevant to the content of this article. Or: Author A has received research
grants from Company W. Author B has received a speaker honorarium from
Company X and owns stock in Company Y. Author C is a member of committee Z.
\end{credits}

\end{comment}
\newpage
\bibliographystyle{splncs04}
\bibliography{refs}




% LNCS Homepage, \url{http://www.springer.com/lncs}, last accessed 2023/10/25
% \end{thebibliography}
\end{document}